\documentclass{emulateapj}

\def\ms{m~s$^{-1}$}

\def\msini{$M_P\sin{i}$}

\def\msun{$M_{\odot}$}
\def\mjup{$M_{\rm Jup}$}

\def\feh{[Fe/H]}

\begin{document}

\title{International Year of Astronomy Invited Review on Exoplanets}    

\author{John Asher Johnson\altaffilmark{1} 
}

\email{johnjohn@ifa.hawaii.edu}

\altaffiltext{1}{Institute for Astronomy, University
  of Hawaii, Honolulu, HI 96822; NSF Astronomy and Astrophysics
  Postdoctoral Fellow} 

\begin{abstract}
Just fourteen years ago the Solar System represented the only known
planetary system in the Galaxy, and conceptions of planet formation
were shaped by this sample of one. Since then, 320 planets have been
discovered orbiting 276 individual stars. This large and growing
ensemble of exoplanets has informed theories of planet formation,
placed the Solar System in a broader context, and revealed many
surprises along the way. In this review I provide an
overview of what has been learned from studies of the occurrence,
orbits and physical structures of planets. After taking a look back
at how far the field has advanced, I will discuss some of the
future directions of exoplanetary science, with an eye toward the
detection and characterization of Earth-like planets around other
stars. 
\end{abstract}

\section{Introduction}

The state of knowledge on planetary systems has undergone a major
revolution over the past 14 years. Starting with the discovery of the
first exoplanet orbiting a normal, hydrogen-burning star in 1995
\citep{mayor95}, the 
sample of known exoplanets has rapidly expanded from a sample of one
to 320 individual planets residing in 276 planetary
systems\footnote{http://exoplanets.org and http://exoplanet.eu}. The
majority of these planets were detected by either Doppler techniques
or by photometric transit surveys, and therefore have
well-characterized orbits with system parameters amenable to uniform
statistical analyses \citep{butler06,torres08}. Additional planets
have been discovered using 
gravitational microlensing and a handful of planets have even been
directly imaged \citep{beaulieu06, gaudi08, kalas08, marois08}. The
occurrence rates, orbital properties, and physical 
characteristics of these worlds inform our understanding of
the formation and orbital evolution of planets in general, and the
origin of our Solar System in particular.  

\section{Planet Occurrence}

The search for exoplanets began with a humble and ancient question: do
planets 
exist around other stars. Hints initially emerged with the detection
of Kuiper Belt-like dust disks around young stars such as Vega
and $\beta$ Pic \citep{aumann84, smith84},  the radial velocity
(RV) detection of 
progressively smaller substellar companions \citep{latham89}, and
the discovery of ``pulsar planets'' \citep{pulsar}. Since then, the
study of planet occurrence 
has evolved from a question of existence to a full-fledged statistical
study of hundreds of systems. Where planets are found and
their relative frequencies around stars of various types provide
valuable insights into the planet formation process and guide future
planet search efforts. 

\begin{figure}[!ht]
\epsscale{1.2}
\plotone{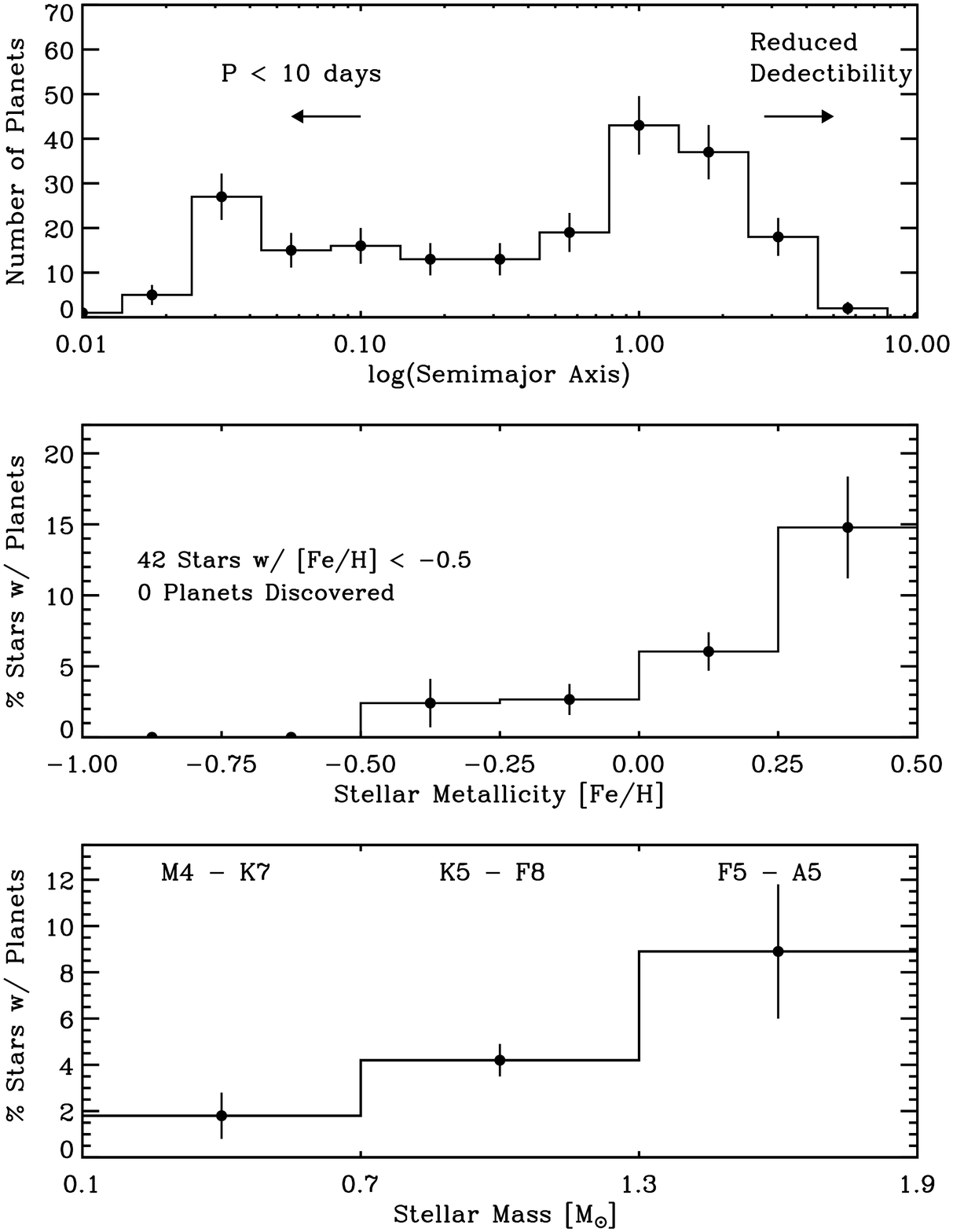}
\caption{{\it Top:} Distribution of Doppler-detected exoplanets as a
  function of 
  semimajor axis. 10\% of stars have planets within 5 AU
  \citep{cumming08}. {\it Middle and Bottom}: The occurrence rate of
  planets rises as a function of metallicity
  \citep{fischer05b} and stellar mass
  \citep{johnson07b}.  \label{plotocc}}     
\end{figure}

Doppler surveys of thousands of stars have shown that among the
Sun-like, FGK stars in the solar neighborhood, roughly one in ten
harbors an giant planet with a period $P < 2000$~days and a minimum
mass greater than half the mass of Jupiter
\citep[\msini~$>0.5$~\mjup\,;][]{cumming08}. The distribution 
of detectable planets as a function of semimajor axis, $dN/d\log{a}$,
is roughly flat 
out to 1~AU, and then rises toward $\sim4.5$~AU, a cutoff corresponding
the decade-long time baselines of the Doppler-based planet
searches (Figure \ref{plotocc}). Careful extrapolation to larger
semimajor axes indicates 
that 17-19\% of stars harbor a giant planet within 20~AU
\citep{cumming08}, and the planets discovered by direct imaging
suggest that planets exist out to semimajor axes of hundreds of AU  
\citep{kalas08, chiang08, veras09}. 

Additional insight has been gained by studying planet occurrence as a
function of stellar properties. The properties of stars are closely
related to the properties of the circumstellar environment during the
early epoch of planet formation. Planet host stars therefore
represent an  important link between the systems detected today and
the processes of planet formation that took place in the past.

For example, studies of the chemical compositions of planet-host stars
reveals a 
strong correlation between stellar metallicity and planet occurrence
\citep{gonzales97, santos04, fischer05b}. The probability of a star
having a detectable planet rises from roughly 3\% around stars with
solar iron abundance (\feh~$ = 0$), to $> 15$\% for stars with \feh~$
> +0.25$ (Figure~\ref{plotocc}; Fischer \& Valenti 2005). The
planet-metallicity correlation is currently best 
understood in the context of the core  accretion theory of planet
formation, in which planets are formed through the collisional buildup
of refractory material in the protoplanetary disk \citep{pollack96,
  ida04b}. A higher stellar metallicity observed today is a reflection
of the higher dust content of the disk while planets were forming.

Mass is another stellar property closely related to the surface
density of solids in the disk midplane, and the core accretion model
predicts that planet frequency should correlate with stellar mass
\citep{laughlin04, ida05a, kennedy08}. This prediction has been
confirmed by Doppler surveys of stars at either end of the mass
scale (Figure~\ref{plotocc}). At the low--mass end, only 4 out of
$\approx 300$ M dwarfs in 
various Doppler surveys have been found to host a Jupiter-mass planet
\citep{marcy98, butler06b, johnson07b, bailey08}. At the other end of
the scale, studies of evolved, intermediat-mass stars ($1.5 < M_*
\lesssim 3$~\msun) on the
subgiant and giant branches \citep{reffert06, sato07, nied07} have
revealed an enhanced planet occurrence 
rate around high-mass stars, rising from $< 2$\% around M
dwarfs, to approximately 9\% around F and A stars \citep{lovis07,
  johnson07b}.   

The correlation between planet occurrence and stellar properties not only
informs theories of planet formation, but also guides the target
selection of future planet searches. The increased ``planeticity'' 
of metal-rich stars has been harnessed by several
metallicity-biased planet searches to find large numbers of
short-period planets around nearby ($d \lesssim 200$~pc) stars
\citep{laughlin00,fischer05a,dasilva06}. The correlation between
stellar mass and 
planet occurrence will be an important consideration for the target
selection of current and future direct-imaging planet search
missions. Indeed, the first imaged planet candidates were discovered 
around the $\sim2$~\msun\ A-type dwarfs Fomalhaut,  HR\,8799 and
$\beta$~Pic \citep{kalas08, marois08, lagrange09}.     
\\
\\

\section{The Observed Physical and Orbital Properties of Exoplanets}

\begin{figure}[!ht]
\epsscale{1.2}
\plotone{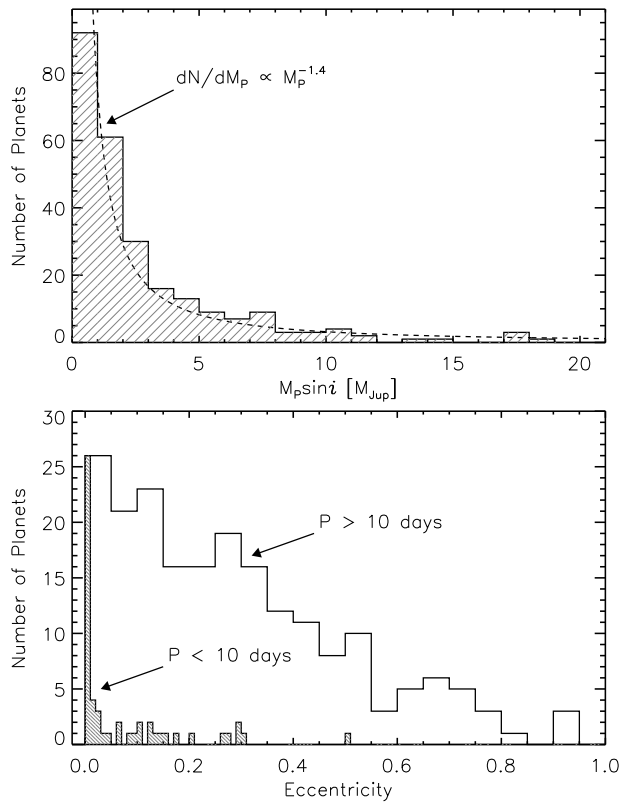}
\caption{{\it Upper:} Distribution of minimum masses (\msini) for all
  Doppler-detected planets. The dashed line shows the best fitting
  power-law. {\it Lower:} Eccentricity distribution of
  Doppler-detected planets for $P < 10$~days (shaded) and $P >
  10$~days (unshaded). \label{plotprops}}   
\end{figure}

Doppler surveys have uncovered a wealth of information about
the orbital characteristics of exoplanets. Giant planets around other
stars have a wide range of semimajor axes and orbital
eccentricities \citep{udry03b, butler06}. In contrast to the nearly
circular orbits of 
the Solar System gas giants, the orbits of exoplanets range from
circular to comet-like, spanning the range $0 \leq e \leq 0.93$, 
with a median eccentricity of 0.24 for $a > 0.1$~AU
(Figure~\ref{plotprops}). The observed eccentricity
distribution is currently best reproduced by simulations of dynamical 
interactions among multiple planets immediately following 
the dissipation of the protoplanetary gas disk \citep[][however, see
  references therein for alternative models.]{juric08, chat08, ford08}.

The distribution of exoplanet minimum masses, \msini, is well fit by the
power-law relationship $dM_P/dN \propto M_P^{-1.4}$ (Figure \ref{plotprops}), 
indicating that smaller planets form more readily than massive
``super-Jupiters'' and brown dwarfs. The paucity of planets with
\msini~$> 10$~\mjup\ is 
known as the ``brown dwarf desert,'' and high-contrast imaging surveys
show that companions in this mass range are rare even out to large
semimajor axes 
\citep{mccarthy04, bddesert}. However, there are indications that the
most massive Jovian planets with \msini~$ \gtrsim 2$~\mjup\ and wide
orbits ($a > 1$~AU) are more prevalent around massive stars
\citep{lovis07, johnson07, sato08a}. 

Doppler surveys have shown that 1.2\% of stars have
planets with orbital periods $P < 10$~days, corresponding to
$a\lesssim0.1$~AU \citep{marcy05a}. These short-period planets are
commonly referred to as "hot Jupiters," and they likely did not form
in situ due to the high temperatures and low surface densities of the
inner regions of protoplanetary disks. Instead, hot Jupiters and other 
close-in planets most likely formed beyond the ``ice line'' at a few
AU, and then migrate inward to their present locations
\citep{lin96, trilling98}. The process of inward migration is now
thought to be a ubiquitous and integral feature of how planets form
\citep{alibert05}.    

\begin{figure}[!h]
\epsscale{1.2}
\plotone{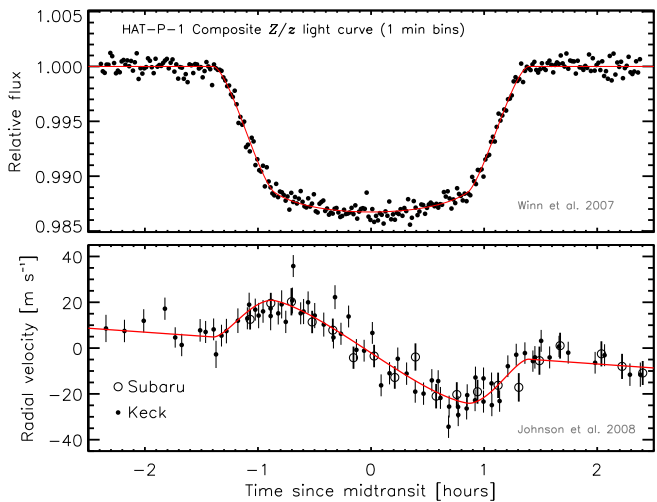}
\caption{Transit observations of the exoplanet HAT-P-1b, which was
  discovered by the HAT-Net wide--field transit survey
  \citep{bakos07}. The planet has a mass $M_P = 0.53$~\mjup, and an 
  orbital period $P = 4.465$~days. {\it Upper:} A composite transit
  light curve constructed from 7 individual photometric data
  sets \citep{winn07b}. The photometry was phased by subtracting the
  mid-transit time from each light curve. The best-fitting model
  gives a planet radius $R_P = 1.225 \pm 0.056 R_{Jup}$ {\it Lower:}
  Phased radial   velocity measurements made during three transits,
  illustrating the 
  Rossiter-McLaughlin (RM) effect. The shape of the
  RM waveform yields the projected spin-obit angle. The HAT-P-1
  planetary system is well aligned with a spin-orbit angle of $3.7 \pm
  2.1$ degrees \citep{johnson08b}. \label{transit}}     
\end{figure}

Planets with orbits that are serendipitously viewed edge--on
transit their host stars and provide valuable additional information
about the physical and orbital characteristics that cannot be studied
by Doppler techniques alone \citep{charbonneau05, winn08r}. The sample of
known exoplanets includes 52 examples of 
well-characterized transiting systems. While the brightest examples ($V
< 8$) come from photometric follow-up of Doppler-detected planets
\citep{henry00b, charbonneau00, bouchy05}, the majority of transiting
planets were detected by wide-field, photometric surveys
\citep{konacki03,tres1,bakos07,wasp1,xo1}. By monitoring hundreds of
thousands  of stars per night using networks of small-aperture,
wide-field 
cameras, these surveys have uncovered a diverse collection of
short-period planets with a wide range of masses, radii and orbital
configurations \citep{southworth08, torres08}. 

The orbit solution of Doppler-detected planets yields the minimum mass
of the planet, \msini, but the light curve of a transiting planet
yields a direct measure of the planet's inclination, providing an
absolute measure of the planet's mass, assuming the stellar mass is
known by other means. The depth of the photometric
dip is related to the planet's 
radius (upper panel of Figure~\ref{transit}). The mass and radius
together give the mean density, which can 
be compared to theoretical models of planetary interiors to provide a
glimpse of their interior structures \citep{sato05, laughlin05b}. 
Modeling efforts of this type have revealed that enhanced heavy
element abundance, ostensibly in the 
form of a large solid core, is a common feature of hot Jupiters
\citep{gillon07, burrows07c, fortney07}. The 
growing sample of transiting planets exhibit heavy-element masses
ranging from 
tens of earth masses up to $\sim100 M_\oplus$. For comparison, Jupiter
is composed of 1--39~$M_\oplus$ of heavy elements, and has a core mass
ranging from 0--11~$M_\oplus$ \citep{saumon04}. Notably, the core
masses of transiting planets correlate with the metal abundances of
the host stars, a finding that lends additional support for the
core accretion model of planet formation
\citep{guillot06,burrows07c,torres08}.  

While many transiting exoplanets have large cores, others have radii
that far exceed the predictions of the current suite of planetary
interior models, even if a heavy-element core is omitted. These
bloated planets pose serious challenges to theoretical models of
planetary interiors \citep{brown01, mandushev07, baraffe03}. The
solution may lie in 
improved  stellar age estimates, revised atmospheric opacities
or a better understanding of chemistry and dynamics in planetary
atmospheres \citep{chabrier04,burrows07b}.  

Planet transits also yield a measure of an
additional fundamental orbital characteristic of planetary systems:
the spin-orbit 
alignment \citep{queloz00}. Measurements of a host star's radial
velocity during the transit of its planet can reveal 
anomalous Doppler-shift variations known as the Rossiter-McLaugnlin
effect \citep{winn05, gaudi07}, and the time-dependent variation of
this effect is related to the 
projected angle between the star's spin axis and the planet's orbit
normal, $\lambda$ (lower panel of Figure~\ref{transit}). The
spin-orbit angle is a fundamental property of planetary systems,
analogous to the  eccentricity and semimajor axis. The majority
of measured exoplanet spin-orbit angles are consistent with $\lambda =
0$, and a single, isotropic distribution for exoplanet spin-orbit
angles can be ruled 
out with high confidence, even after accounting for projection effects
\citep{winn05, fabrycky09}. This result suggests that the dominant
mechanism responsible for the inward migration of planets preserves
spin-orbit alignment (although \citet{hebrard08} and \citet{winn09b}
present a notable exception in the misaligned XO-3 system). 

Half an orbital period after a planet transit, another
observational opportunity arises when the planet passes behind the
parent star. Observations made during these occultation
events provide a means of detecting light from the planet itself, as
revealed by the flux decrement caused by the star blocking the light
from the planet \citep{demming05, charbonneau05}. The variation in the
depth of the occultation light 
curve measured in different bandpasses provides a measure of the
planet's emission spectrum, which can then be compared to theoretical
models to gain insights into the characteristics of planetary
atmospheres. Observations of this type made with space-based
facilities such as {\it Spitzer Space Telescope}, the {\it Hubble
  Space Telescope}  and MOST, have revealed evidence of temperature
inversions in the atmospheres of some planets \citep{knutson09};
detected molecules such as H$_2$0 and CO$_2$ in their atmospheres
\citep{grillmair08};  placed limits on their wavelength-dependent bond  
albedos \citep{rowe08}; and have been used to measure their
eccentricities based on the time interval between primary and
secondary eclipse \citep{demory07}. 

\section{Multiplicity}

\begin{figure}[!h]
\epsscale{1.1}
\plotone{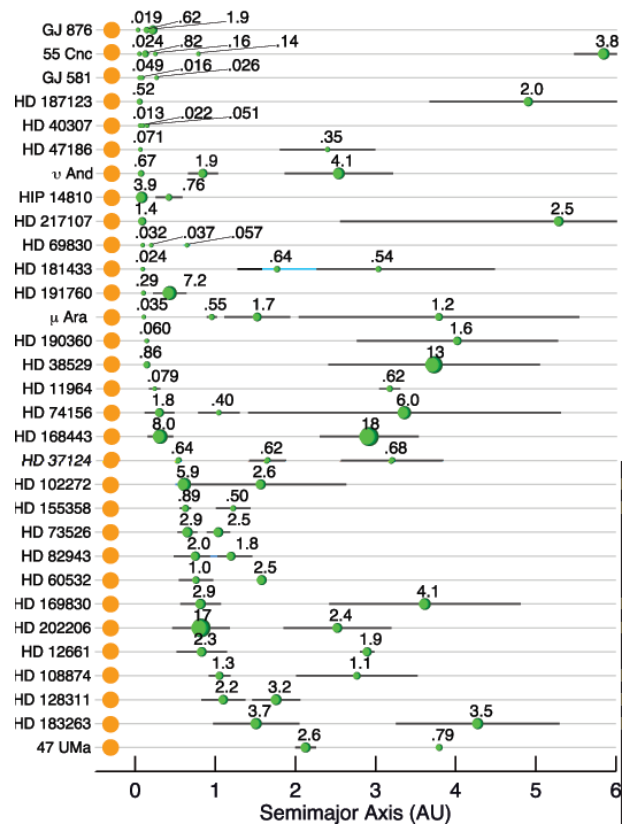}
\caption{Chart showing the semimajor axes and minimum mases for the 31
  known planetary systems containing more than 1 planet
  (data from Wright et al. [2009], as updated by Correia et
  al. [2009], Mayor et al. [2009] and \citet{nied09}). The 
  horizontal lines under each planet indicate the distance between
  periapse and apoapse for eccentric orbits. The sizes of the
  planet symbols scale as (\msini)$^{1/3}$, and the labels denote the
  planet mass in units of \mjup. \label{plotmulti}}   
\end{figure}

There are 31 known multi-planet systems orbiting nearby stars with
well-characterized orbits \citep[Figure
  \ref{plotmulti};][]{wright09, udry07}. These systems, together 
with single-planet 
systems with additional radial velocity trends, comprise
30\% of all known planetary systems within 200~pc. Five of the known
multi-planet systems systems are in mean-motion resonances, and two 
are known to contain four or more planets: the system of five planets 
around 55 Cnc and the four-planet system around
$\mu$~Ara \citep{fischer08, pepe07}. The characterization of
multi-planet systems requires 
intensive follow-up observations, and the additional scrutiny has
resulted in the detection of some of the least massive planets currently
known. The Gl876, 55~Cnc and $\mu$~Ara systems all contain low-mass planets
that were discovered by intensive Doppler follow-up, with \msini~$ =
5.89$, 10.8 and 15~$M_\oplus$, respectively
\citep{rivera05, mcarthur04, fischer08, pepe07}. The discovery of
multi-planet systems has 
been aided by improvements to the attainable Doppler precision,
resulting in planetary systems containing multiple Neptune- and
sub-Neptune-mass planets \citep{vogt05, lovis06, mayor09}. The
increasing time baselines of Doppler surveys, together with the
sensitivity of microlensing surveys, are bringing Solar System analogs
within reach \citep{wright07, gaudi08}.  

Multi-planet systems are valuable testing grounds for theories of
planet formation and migration \citep{ford06r,terquem07}. For
example, planets should form at arbitrary locations 
within the protoplanetary disk, and yet several multi-planet systems
have been detected in mean-motion resonances. These systems indicate
that systems 
of planets migrate inward and are subsequently captured into 
resonant configurations \citep{marcy01, correia09,
  raymond08}. Comparisons between the properties of multi-planet systems
and those of systems containing only a single detectable planet
provide important observational constraints for theories of
planet-planet and disk-planet interactions \citep{alibert05,
  ford05b}. 
\\

\section{Future Directions}

The low-mass planets in short-period orbits detected by Doppler-based
planet searches \citep{mayor09} and those at wider separations detected via 
gravitational microlensing \citep{bennett08} point the way toward the
future of exoplanet science: the detection and characterization of
Earth-like planets. One of the primary goals of the study of
exoplanets is to determine whether solar systems like our own exist
elsewhere in the Galaxy \citep{taskforce}. However,
the detection of an Earth-mass planet in the terrestrial zone of a
Sun-like star (late F-type to early K) poses a considerable technical
challenge. These small planets 
induce radial velocity variations of $\sim10$~cm~s$^{-1}$, astrometric
variations of a $\sim 0.1$~microacrseconds ($\mu$as), transit depths
of $\sim 10^{-4}$, and have planet-star contrast ratios of $10^{-10}$
to $10^{-7}$, for reflected light and thermal emission, respectively
\citep[e.g.][]{schneider02}.  

One immediate method of circumventing these technical
hurdles is to first search for habitable worlds around M dwarfs. The
amplitude of the Doppler variations induced by a planet of a given
mass in the habitable zone of a star scale as $K \propto
M_*^{-3/2}$, and the transit depth scales as $M_*^{-2}$ (assuming $R_*
\propto M_*$ for he lower main sequence).
The detection sensitivities are therefore an order of magnitude higher
around a 0.25\msun\ M dwarf compared to a solar-type star. 
Targeted photometric and Doppler searches for transiting terrestrial
planets in the habitable zone of M dwarfs are currently underway
\citep{butler04, bonfils05b, irwin08}. The 
low-mass planetary system tantalizingly close to the habitable zone of
the  0.4~\msun\ M Dwarf Gl581 demonstrates the promise of searching
for planets around low-mass stars \citep{udry07}.

The occurrence rate of terrestrial planets around Sun-like stars can
be estimated  from the frequency of exoplanets with masses intermediate
to Earth and Neptune, commonly known as ``super-Earths.'' Planets with
masses $\sim 10 M_\oplus$ in short-period orbits are readily
detectable by Doppler surveys capable of $\sim 1$~\ms\ precision, and
several dedicated searches are currently underway using HARPS and
Keck/HIRES \citep{mayor09, howard09}. While these planets have masses
approaching the Earth's, it remains unclear whether they are
Earth-like with solid surfaces, or instead miniature Neptunes with
massive envelopes of H and He \citep{valencia07, barnes09}. This issue
will soon be addressed by the detection of transiting super-Earths
around nearby, bright ($ V < 9$) stars, for which Doppler follow-up
can provide precise mass measurements. The 
recent demonstrations of sub-millimagnitude, ground-based transit
photometry, together with the growing
number of low-mass planets in short-period orbits suggest that the
first nearby ($d \lesssim 50$~pc) transiting super-Earth will be
detected in the near future \citep{johnson09, gillon09, winn09}.   

Searching for Earth-analogs around Sun-like
stars is best pursued above the Earth's atmosphere. The COROT
and {\it Kepler} space   
missions will provide continuous high-precision photometric monitoring
of hundreds of thousands of stars within select patches of the
sky to detect planets with a wide range of masses and orbital
characteristics. Early results from COROT demonstrate the advantages
afforded by observing from space, both in terms
of attainable photometric precision and continuous temporal coverage,
which increases detection sensitivity at longer orbital
periods compared to ground-based surveys \citep{aigrain08}. The {\it
  Kepler} mission launched successfully in March of this year and 
will maintain a photometric precision of 20 micro-magnitudes over the
course of 4 years \citep{kepler}. The precision and time baseline of
{\it Kepler} will provide a galactic census of terrestrial planets 
around Sun-like stars out to separations of 1--2 AU. The transiting
systems detected by {\it Kepler} and COROT will open up exciting new
science directions for the next generation space observatories such as
the James Webb Space Telescope (JWST), including atmospheric
transmission spectroscopy and measuring the phase variations of
thermal and reflected light \citep{seager08}.

The ``holy grail'' is the
imaging detection, and subsequent spectroscopic study, of a
terrestrial planet in the habitable zone of a nearby star. Space-borne
astrometry 
will provide a means of detecting and directly measuring the masses
and orbital configurations of terrestrial planets. One such mission on
the near horizon is the {\it Space Interferometry Mission}
(SIM--Lite), which will provide an astrometric precision of better
than 1~$\mu$as \citep{simlite}. Once Earth-like planets are
identified, high-contrast imaging using 
techniques such as adaptive 
optics and coronography can be brought to
bear to measure colors, and possibly even spectra, to search for
biosignatures. Ground-based imaging
surveys are making impressive strides, and planet searches are now
beginning with the NICI campaign \citep[][; Liu et
  al. 2009 in press]{nici} and 
in the near future with the Giant Planet Imager \citep{gpi}, with the
goal of detecting Jupiters in wide orbits. The technology  developed
for and proven by these surveys, and others like them, will 
inform future imaging efforts from space, such as the {\it Terrestrial
  Planet Finder} \citep{tpf}.

In just 14 years conceptions of planets around other stars have
evolved from science fiction to a mature field of scientific
inquiry. The next decade holds much promise as we progress toward the
discoveries of solar systems like our own around other stars.

\acknowledgements 
I gratefully acknowledge Geoff Marcy, Andrew Howard, Debra Fischer,
Jason Wright and Michael Liu for their helpful conversations, feedback
and edits. I am an NSF Astronomy and Astrophysics Postdoctoral Fellow
with support from the NSF grant AST-0702821.

\end{document}